\documentclass[epj]{webofc}
\usepackage[varg]{txfonts}   
\usepackage{hyperref}
\usepackage{textcomp}
\usepackage{color}

\renewcommand*{\eqref}[1]{Eq.~(\ref{eq:#1})}

\wocname{\includegraphics[width=0.25cm,clip]{logoARENA} ARENA2018}
\wocname{ARENA2018}
\woctitle{ARENA2018}
\woctitle{\includegraphics[width=0.25cm,clip]{logoARENA} ARENA2018}
\begin{document}
\title{Physics Potential of a Radio Surface Array at the South Pole}
%
%

\author{\firstname{Frank G.} \lastname{Schr\"oder}\inst{1,2}\fnsep\thanks{\email{frank.schroeder@kit.edu} ~~and ~~\email{fgs@udel.edu}}
for the IceCube-Gen2 Collaboration}


\institute{Institute of Experimental Particle Physics, Karlsruhe Institute of Technology (KIT), Karlsruhe, Germany  
\and
           Department of Physics and Astronomy, University of Delaware, Newark, DE, USA
          }

\abstract{%
A surface array of radio antennas will enhance the performance of the IceTop array and enable new, complementary science goals. 
First, the accuracy for cosmic-ray air showers will be increased since the radio array provides a calorimetric measurement of the electromagnetic component and is sensitive to the position of the shower maximum. 
This enhanced accuracy can be used to better measure the mass composition, to search for possible mass-dependent anisotropies in the arrival directions of cosmic rays, and for more thorough tests of hadronic interaction models. 
Second, the sensitivity of the radio array to inclined showers will increase the sky coverage for cosmic-ray measurements.
Third, the radio array can be used to search for PeV photons from the Galactic Center. 
Since IceTop is planned to be enhanced by a scintillator array in the near future, a radio extension sharing the same infrastructure can be installed with minimal additional effort and excellent scientific prospects. 
The combination of ice-Cherenkov, scintillation, and radio detectors at IceCube will provide unprecedented accuracy for the study of high-energy Galactic cosmic rays.
}
\maketitle
\section{Introduction: Present Situation}
\label{intro}
Cosmic-ray science in the energy range up to a few EeV ($= 10^{18}\,$eV) and the search for high-energy photons are important science goals of IceCube after its central purpose as a detector of high-energy neutrinos.
The IceCube experiment detects air showers by a dedicated array of ice-Cherenkov detectors at the surface, called IceTop \cite{IceTopNIM2013}. 
Additionally, the IceCube in-ice array of photomultipliers, built primarily to detect neutrinos, tracks high-energy muons produced in air showers. 
IceTop provides measurements with high statistics for cosmic rays from a few $10^{14}\,$eV onwards. 
Due to its high altitude of $2800\,$m it measures close to the maximum of nearly vertical showers and thus has a high accuracy for the reconstructed energy of the primary particle. 
The combination of in-ice muons with the surface signal provides a mass estimator for nearly vertical air showers.
However, the interpretation of these measurements suffers from systematic uncertainties, because Monte Carlo simulations of air showers have known deficiencies regarding the muonic component \cite{2017ApelKG_MuonAttenuation}, \cite{AugerMuonNumber2015}. 
In contrast to other air-shower experiments, IceCube/IceTop does not yet feature optical or radio detectors providing complementary measurements of $X_\mathrm{max}$, the atmospheric depth of the shower maximum. 
Moreover, the sensitivity of IceTop to the electromagnetic component is continually degraded by snow accumulation.

\begin{figure}[b]
 \centering
  \includegraphics[width=0.9\textwidth]{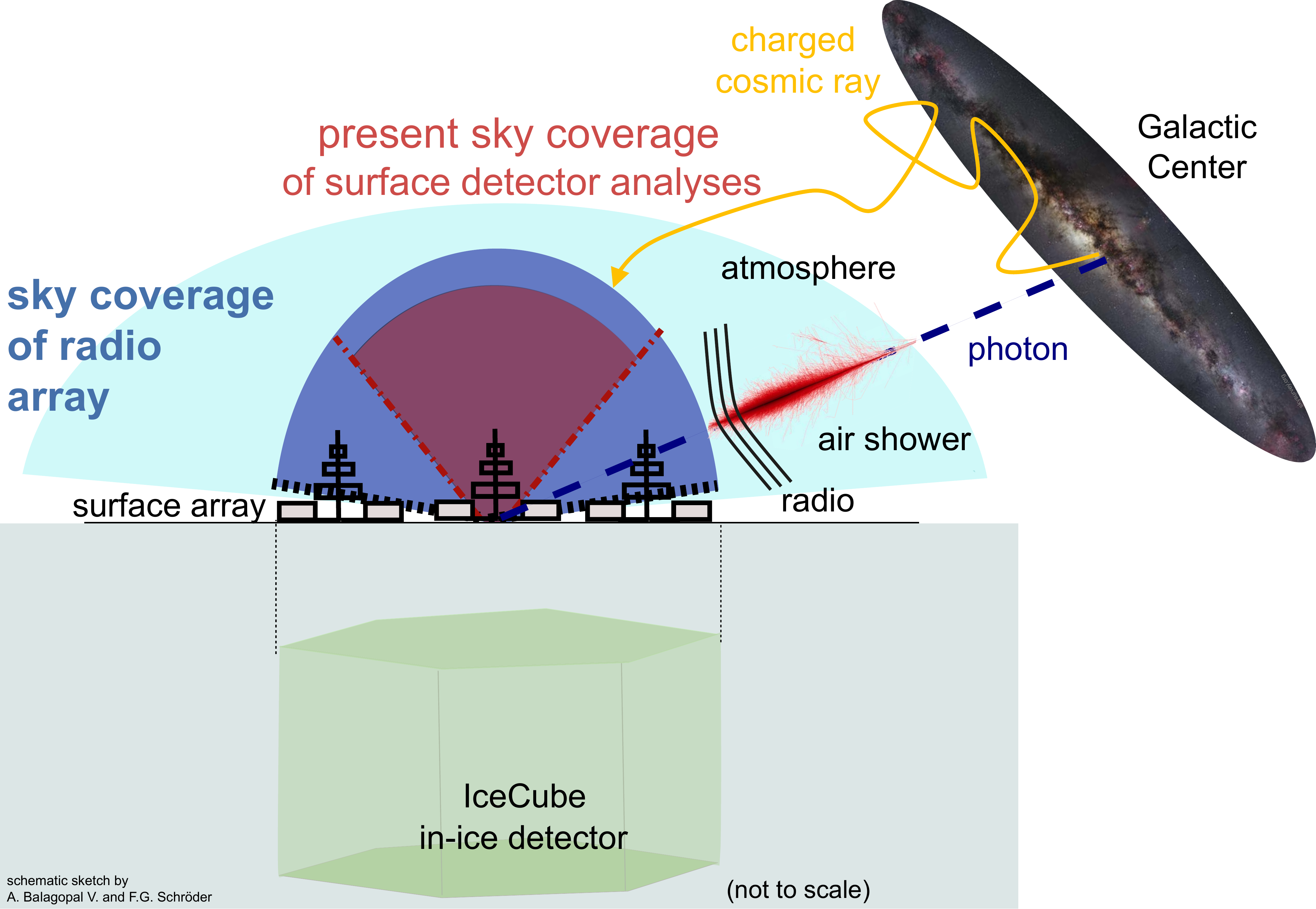}
  \caption{Schematic sketch of a potential radio extension of the IceCube surface array. The radio array will improve the accuracy and extend the field of view to lower elevations. 
  In addition to cosmic-ray measurements, the radio array can search for photons of PeV energy from the Galactic Center.
  }
  \label{fig_sketch}
\end{figure}   

The sensitivity to the electromagnetic component will be achieved for zenith angles $\theta \lesssim 40^\circ$ by the installation of a scintillator array at the surface \cite{IceScint_ICRC2017}. 
This array enable the calibration of the effect of the snow coverage. 
Furthermore, it will enhance the mass sensitivity of IceTop by measuring the electron-muon ratio in combination with the ice-Cherenkov tanks. 
However, systematic uncertainties due to the limitations of the hadronic interaction models will remain. 
To use the full potential of IceCube for cosmic-ray science, an additional detector for the electromagnetic shower component is required. 
While the classic techniques of fluorescence and atmospheric Cherenkov light are restricted in duty cycle (at the South Pole to periods of clear weather during the winter months), the radio array can operate continuously and in coincidence with any of the particle detectors.

\section{Scientific Potential of a Radio Extension of the Surface Array} 
Given the recent progress of the radio technique for measurement of air showers \cite{SchroederReview2016, HuegeReview2016}, a surface array of antennas at the South Pole is ideal for enhancing the performance of IceCube for cosmic-ray detection. 
The radio array will provide a calorimetric measurement of the electromagnetic shower component and is sensitive to the position of its maximum. 
If triggered by the particle detector arrays, it will improve the total measurement accuracy for showers above its energy threshold. 
Additionally, the radio array can measure inclined showers for which IceTop alone as well as the future scintillator array have a poor reconstruction accuracy. 
Therefore, the sky coverage for all types of cosmic-ray analyses will be significantly enhanced by the radio array. 
The combination of improved accuracy and sky coverage will enhance the performance of IceCube for several of its present science goals and enable it to study new aspects:

\noindent \textbf{Energy Spectrum and Mass Composition}\\
The flux and mass composition of the primary cosmic rays as a function of the energy are classical measures in cosmic-ray science. 
Features in these observables carry information on the propagation and the origin of the cosmic rays. 
Due to the lower systematic uncertainties of the radio technique, the surface antenna array will increase the per-event precision and the absolute accuracy. 
Finally, the energy scale of IceCube can be compared to other cosmic-ray experiments with a radio extension \cite{TunkaRexScale2016}.
\\

\noindent \textbf{Mass-dependent Anisotropy}\\
Weak anisotropies in the all-particle flux of primary cosmic rays have been discovered below $2\,$PeV \cite{IceCubeAnisotropy2016} and above $8\,$EeV \cite{AugerScience2017}. 
The maximum energy of cosmic rays produced in the Milky Way is presumed to be in this energy range.
However, the arrival directions in this energy range are remarkably isotropic, possibly because of the overlaping of different anisotropies of Galactic and extragalactic cosmic rays. 
Since extragalactic cosmic rays below $10^{19}\,$eV seem to be mainly protons or alpha particles \cite{AugerCombinedFit2017}, \cite{AugerTAcomparison_ICRC2017}, but Galactic cosmic rays at these energies have a significantly heavier composition \cite{KGheavyKnee2011}, \cite{IceTopCompositionMethod_ICRC2017}, to investigate this requires an event-to-event separation of light and heavy cosmic rays. 
While the scintillator extension will provide this separation for nearly vertical showers, the radio extension will provide it for more inclined events.
\\
 
\noindent \textbf{Particle Physics in Air Showers}\\
The separate measurement of the electromagnetic component by the radio array, of the electrons and low-energy muons by the particle detectors at the surface, and of the high-energy muons in the ice will enable more thorough tests of hadronic interaction models. 
Improving the knowledge of air-shower physics at multi PeV energies, and in particular understanding the 'muon deficit' in the models, is necessary to better quantify the muon background in neutrino measurements by IceCube.
\\

\noindent \textbf{PeV Photon Search}\\
Photon induced showers emit a slightly stronger radio signal than showers initiated by cosmic-ray nuclei and contain an order of magnitude fewer muons. 
Thus, the combination of radio and muon detectors is ideal to search for photons. 
The Galactic Center, a promising candidate source of PeV photons \cite{Balagopal2018}, is continuously visible at $61^\circ$ zenith from the South Pole, thus providing significant discovery potential.
\\

\noindent \textbf{Pathfinder for the Next Generation}\\
In addition to its primary goals, a radio surface array will serve as a pathfinder for next generation projects. 
In particular, it can be used to study the potential of a large-scale array of surface antennas in combination with other extensions of IceCube-Gen2, e.g., for calibration of an in-ice radio array. 
Due to the large radio footprint of inclined showers \cite{AERAinclinedJCAP2018}, a radio surface array could veto very inclined events, even if the shower core is far outside the array. 
Furthermore, by using antennas developed for the Square Kilometer Array (SKA), a radio extension of IceTop will become its natural pathfinder. 
The SKA has a complementary science case, since it will measure the radio signal of air showers in great detail \cite{Zilles_ARENA2016_SKA}, but will not feature muon detectors.


\begin{figure}[t]
 \centering
  \includegraphics[width=0.57\textwidth]{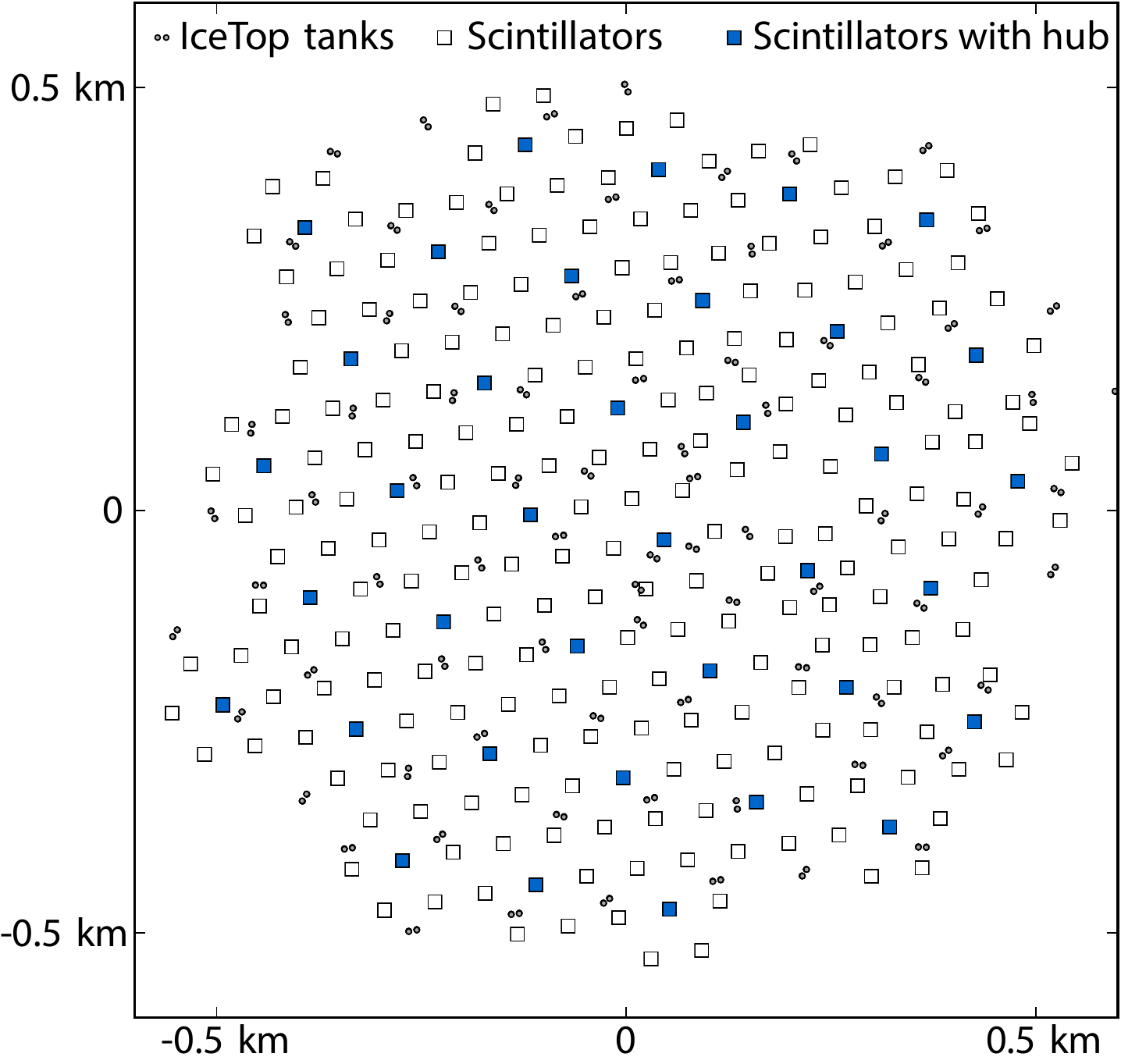}
  \hfill 
  \includegraphics[width=0.36\textwidth]{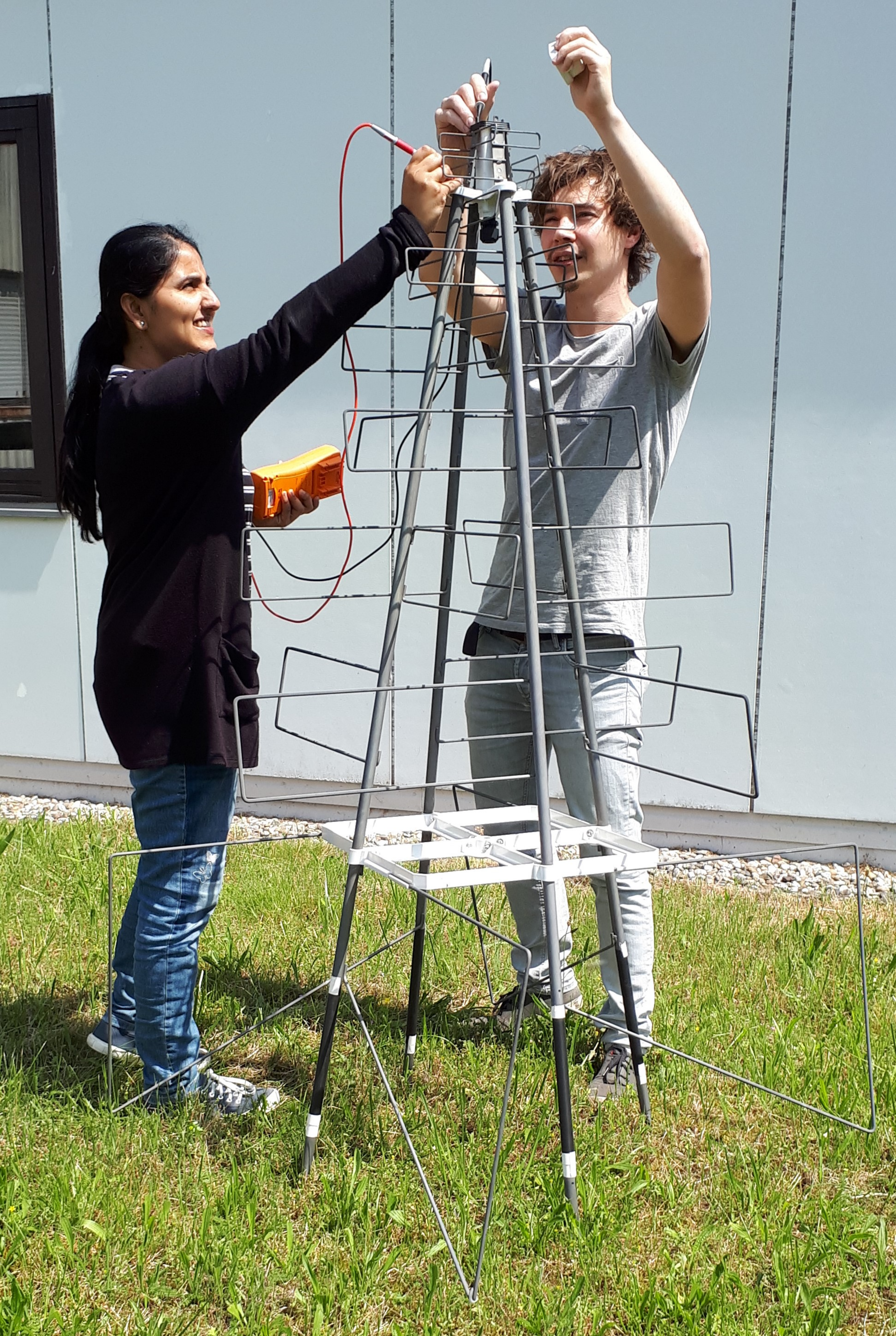}
  \caption{Left: Map of the particle-detector arrays at the surface of IceCube; the IceTop array in operation since several years, and the proposed scintillator array. 
  Right: SKA prototype antenna used for first tests at KIT.
  }
  \vspace{-0.4cm}
  \label{fig_map}
\end{figure}   

\section{Performance}
The performance of the radio array can be estimated based on other antenna arrays successfully measuring cosmic-ray air showers. 
Moreover, a dedicated study with CoREAS simulations has been performed to determine the optimum frequency band and the energy threshold \cite{Balagopal2018}. 
Since the electromagnetic shower component responsible for the radio emission is similar for all primary particles, these results can be use to estimate the general performance for cosmic rays. 
With present simulations, many key parameters were estimated approximately.
Prototype measurements and further simulation studies are needed to optimize the configuration of the final array. 
\\

\noindent \textbf{Threshold: few PeV}\\
At a radio quiet site the most relevant backgrounds are thermal and Galactic radio noise.
This was confirmed for the South Pole by earlier measurements at lower frequencies using RASTA surface antennas at the ARA testbed \cite{Boeser_RASTA_ARENA2012}, \cite{ARAtestbed2012}.
In addition there is human-made temporary RFI at the South Pole that will be easy to filter for externally triggered radio measurements. 
According to our simulation study assuming a thermal noise of $300\,$K, the optimum band for detection of air showers is around $100-190\,$MHz.
Using the detection criterion of AERA \cite{AugerAERAenergy2015} and Tunka-Rex \cite{TunkaRex_NIM_2015}, i.e., a signal-to-noise ratio of ten in at least three antennas, the threshold will depend on the arrival direction of the shower and is expected of order a few PeV.
\\

\noindent \textbf{Aperture: $0^\circ$ to approximately $70^\circ$ zenith angle}\\
For vertical showers the radio footprint is smallest and will have a diameter of order $250\,$m. 
Full sky coverage requires an antenna spacing dense enough to cover at least three antennas in any possible circle of this diameter.
Hence, around 80 antennas within the $1\,$km$^2$ area of IceTop are required for full sky coverage including vertical showers. 
\\

\noindent \textbf{Primary particle: direction, energy, mass}\\
Since the frequency band of the proposed extension of IceTop will be wider than that of current radio arrays, the reconstruction accuracy is expected to be at least as good as that of similarly sparse arrays. 
LOPES achieved better than $0.5^\circ$ direction accuracy with a timing precision of about $1\,$ns \cite{SchroederTimeCalibration2010}, and an array diameter of only $200\,$m \cite{2014ApelLOPES_wavefront}. 
Tunka-Rex and AERA feature an energy accuracy of the order of $15\,\%$ \cite{AERAenergyPRL2015}, \cite{TunkaRex_Xmax2016}, which can probably be improved down to $10\,\%$ (per-event precision and absolute scale). 
The primary mass can be statistically reconstructed by the measurement of $X_\mathrm{max}$, and by the ratio of the radio amplitude and the muon number.
Interpolating between LOFAR and Tunka-Rex results \cite{BuitinkLOFAR_Xmax2014}, \cite{TunkaRexPRD2018}, the $X_\mathrm{max}$ precision will be around $20-30\,$g/cm$^2$, which corresponds to the average difference between showers initiated by protons and helium nuclei. 
According to the simulation studies performed for the Pierre Auger Observatory \cite{Holt_ARENA2018}, the radio-muon combination is at least as sensitive to the mass composition and provides a complementary method. 
Being the first experiment combining both methods for the multi-PeV energy range will provide unprecedented accuracy for high-energy Galactic cosmic rays.
\\

\begin{figure}[t]
 \centering
  \includegraphics[width=0.8\textwidth]{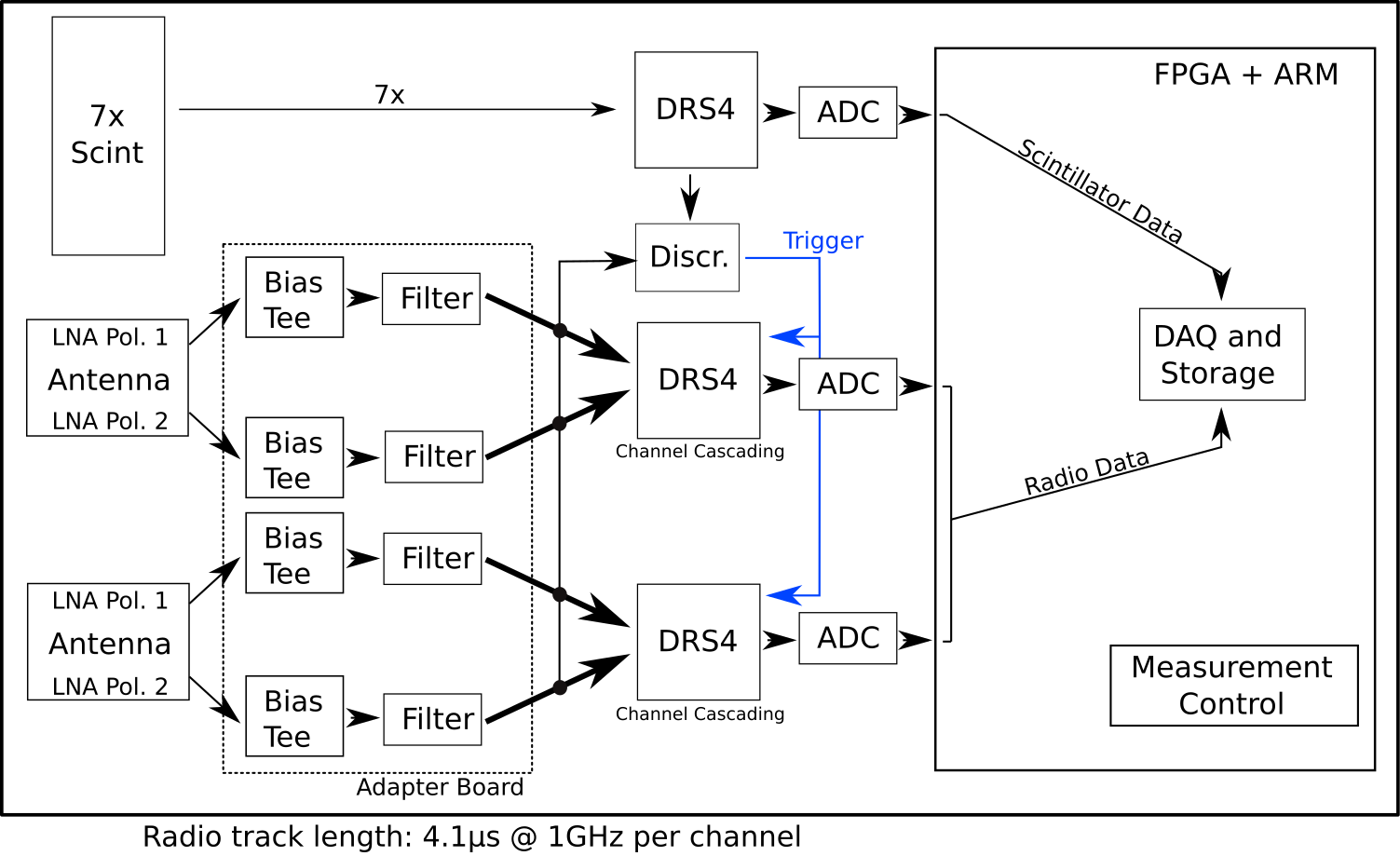}
  \caption{Schematics of the TAXI electronis for the data-acquisition. 
  The TAXI system presently recording data of seven scintillators running at IceTop can easily be adapted to record radio data in addition. 
  In a first prototype two antennas with two polarizations each will be attached. 
  }
  \vspace{-0.4cm}
  \label{fig_TAXI}
\end{figure}

\section{Ideas of the Realization of the Radio Array}
The radio array can be realized in a cost-effective manner by attaching antennas to the planned scintillator array (Fig.~\ref{fig_map}), sharing the infrastructure and data-acquisition. 
Currently, two options for the scintillator design and its data-acquisition electronics are under test at the South Pole, where the first air showers have already been detected. 
By connecting the antennas to the same electronics, their readout can easily be triggered by the scintillators (for judging the potential of self-triggering, the situation concerning temporary RFI has to be studied with prototype antennas). 
One of the two electronics options is the TAXI system using three DRS4 chips for digitization of any signal \cite{Karg_TAXI_ARENA2014}. 
Since only one DRS4 chip is needed for each station of seven scintillators, the remaining two chips can be used to connect the antennas (Fig.~\ref{fig_TAXI}). 
A good candidate for the antenna, is the SKALA antenna developed for the low-frequency array of the SKA telescope. 
SKALA features a low system noise and a smooth gain pattern with a high sensitivity for the complete zenith-angle range of interest \cite{SKALAantenna_v1}, \cite{SKALAantenna_v2}. 
First tests at KIT show that the integrated low-noise amplifier operates well at the low temperatures expected at the South Pole. 
Thus, we propose to install two SKALA antennas soon at the South Pole to perform first prototype measurements triggered by the already operating scintillator prototypes.

\section{Conclusion}
Thanks to the demonstrated progress in the radio technique, and given the window of opportunity provided by the installation of the scintillator array, the time is ripe for a radio extension of IceTop.
Despite the modest resources needed for adding antennas to the scintillator array, the physics potential of the radio surface array is bright. 
It has breakthrough potential in the search for PeV photons from the Galactic Center and the search for mass-dependent anisotropies, along with significant progress in standard cosmic-ray physics.
The expected accuracy and sky coverage will make the combination of IceTop, the new scintillators and the antennas an outstanding cosmic-ray detector.

\bibliography{arena2018} 

\end{document}